\documentclass{raa}

\usepackage{natbib}
\usepackage{graphicx,times}             
\usepackage{amsmath}                
\usepackage{amsfonts}               
\usepackage{amssymb}                
\usepackage{hyperref}

\def \s{~\rm{s}}
\def \km{~\rm{km}}

\def \AU{~\rm{AU}}
\def \erg{~\rm{erg}}

\def \yr{~\rm{yr}}

\def \days{~\rm{days}}

\begin{document}

\title{Operation of the jet feedback mechanism (JFM) in intermediate luminosity optical transients (ILOTs)}

   \volnopage{Vol.0 (200x) No.0, 000--000}      
   \setcounter{page}{1}          

\author{Amit Kashi
      \inst{1}
\and 
      Noam Soker
      \inst{2}
}

\institute{Minnesota Institute for Astrophysics, University of Minnesota, 116 Church St. SE. Minneapolis, MN 55455, USA \email{kashi@astro.umn.edu}
\and
Department of Physics, Technion -- Israel Institute of Technology, Haifa 32000 Israel
}

\abstract{
We follow the premise that most intermediate luminosity optical
transients (ILOTs) are powered by rapid mass accretion onto a
main sequence star, and study the effects of jets launched by
an accretion disk. The disk is formed due to large specific
angular momentum of the accreted mass. The two opposite jets
might expel some of the mass from the reservoir of gas that
feeds the disk, and therefore reduces and shortens the mass
accretion process. We argue that by this process ILOTs limit
their luminosity and might even shut themselves off in this
negative jet feedback mechanism (JFM). The group of ILOTs is a
new member of a large family of astrophysical objects whose
activity is regulated by the operation of the JFM.
}

   \authorrunning{A. Kashi \& N. Soker }            
   \titlerunning{The JFM in ILOTs}  

   \maketitle

\section{INTRODUCTION}
 \label{sec:intro}

The heterogenous group (\citealt{Kasliwal2011}) of eruptive stars with
peak luminosity below those of supernovae and above those of novae
(e.g. \citealt{Mouldetal1990, Rau2007, Ofek2008, Prieto2009,
Botticella2009, Smithetal2009, Berger2009b, KulkarniKasliwal2009,
Mason2010, Pastorello2010, Kasliwaletal2011, Tylendaetal2013,
Kasliwal2013}) has been growing in recent years
(\citealt{Kasliwal2013}). Excluding low luminosity supernovae and
similar objects, such as Ca-rich transients and .Ia SNe, the rest
of the eruptive events are part of a group of the so called
intermediate luminosity optical transients (ILOTs;
\citealt{Berger2009b}).

ILOT events of systems that harbor asymptotic giant branch (AGB)
or extreme-AGB (ExAGB) pre-outburst stars, like NGC~300~OT2008-1
(NGC~300OT; \citealt{Monard2008, Bond2009, Berger2009b}) and
SN~2008S (\citealt{ArbourBoles2008}), were studied both in the frame
of single star models (e.g., \citealt{Thompsonetal2009,
Kochanek2011}) and binary stellar models (\citealt{Kashietal2010,
KashiSoker2010b, SokerKashi2011, SokerKashi2012, SokerKashi2013}).
Recent observations cast doubt that the projenitor star survived these events (\citealt{Adamsetal2015}).
\cite{McleySoker2014} conclude that single-star models for ILOTs
of evolved giant stars encounter severe difficulties, and that
ILOTs are most likely powered by a binary interaction. We here
continue with developing and exploring the binary model for
\emph{all} ILOTs as we outlined in an earlier paper
(\citealt{KashiSoker2010b}).

In \cite{KashiSoker2010b} we developed the High-Accretion-Powered
ILOT (hereafter, HAPI) model. The HAPI model suggests that many
(most) ILOTs are powered by a high-accretion rate event onto a
main sequence (MS), or slightly involved off the MS, star in a
binary system. In some cases the binary system is actually a
triple star system, where the tertiary star induces orbital
instabilities and causes the two inner stars to interact. In this
mass transfer event the accreted mass possesses high specific angular
momentum to form an accretion disk, or an accretion belt, around
the MS star (\citealt{KashiSoker2009}). The accretion disk can launch
two opposite jets that expel more mass from the system to form an
expanding bipolar nebula (\citealt{KashiSoker2010a}), such as the
bipolar nebula of $\eta$ Car, the Homunculus, that was formed in
the Great Eruption (GE; e.g., \citealt{HumphreysMartin2012} and
references therein). The connection between some ILOTs and bipolar
planetary nebulae (PNe) was mentioned in the past
(\citealt{SokerKashi2012, AkashiSoker2013}) in regards to the binary
model for ExAGB-ILOTs. \cite{Prieto2009} already made a connection
between NGC~300OT and pre-PNe. The connection between some bipolar
PNe and and some ILOTs has gained support in recent years (e.g.,
\citealt{BoumisMeaburn2013, Clyneetal2014, DeMarco2014,
DeMarco2015a, DeMarco2015b, Zijlstra2015}).

In this paper we update the classification and grouping of ILOTs
(section \ref{sec:ILOTs}) to further emphasize the common
properties they share, including the presence of jets. In section
\ref{sec:JFM} we turn to propose that in some ILOTs the jets lead
to a negative jet feedback mechanism (JFM). We compare some
properties of the JFM in ILOTs to those in other systems where the
JFM operates. Our short summary is in section \ref{sec:summary}.

\section{ILOTs}
 \label{sec:ILOTs}
\subsection{The Energy-Time Diagram (ETD)}
 \label{subsec:ETD}

The objects studied here (gap objects that are not supernovae) can
be classified in the following way.
\newline
\textbf{ILRT: Intermediate-Luminous Red Transients}. These are
events resulting from evolved stars, such as AGB and ExAGB stars,
and similar objects, e.g., red giant branch (RGB) stars. Most
likely a companion accretes mass and the gravitational energy
powers the eruption, e.g., NGC~300~OT. A similar process occurs in
giant eruptions of luminous blue variables (LBVs).
\newline
\textbf{LBV giant eruptions and SN Impostors}. Giant eruptions of
Luminous Blue Variables. Examples include the 1837--1856 GE of
$\eta$ Car and the pre-explosion eruptions of
SN~2009ip. ILRTs are the low mass relatives of LBV giant
eruptions.
\newline
\textbf{LRN or RT: Luminous Red Novae or Red Transients or
Merger-bursts.} These are powered by a full merger of two stars.
The process of destruction of the less dense star onto the denser
star releases gravitational energy that powers the transient.
Examples include V838 Mon and V1309~Sco. Merger events of stars
with sub-stellar objects are also included.
\newline
\textbf{ILOT: Intermediate-Luminosity Optical Transients.} ILOT is
the term for the combined three groups listed above (ILRT, LRN and
LBV giant eruptions). These events, we argue, share many common
physical processes, in particular being powered by gravitational
energy released in a high-accretion rate event, the HAPI model.

We emphasize again that the definition of ILOT does not include
low luminosity supernovae and similar objects, such as Ca-rich
transients and .Ia SNe (for these see \citealt{Kasliwal2013}).

\begin{figure*}[!t]
\includegraphics[trim= 6.0cm 0.0cm 0.0cm 0.0cm,clip=true,width=0.99\textwidth]{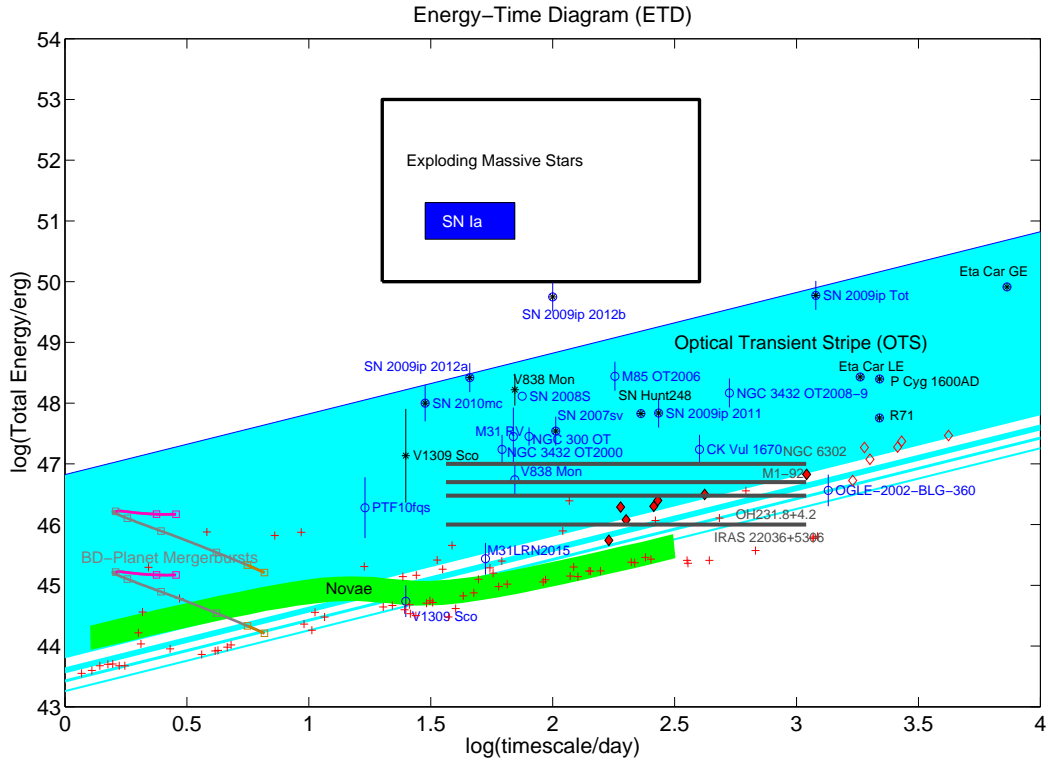}
\caption{{ Observed transient events on the Energy-Time Diagram
(ETD). Blue empty circles represent the total (radiated plus
kinetic) energy of the observed transients as a function of the
duration $t$ of their eruptions. The Optical Transient Stripe
(OTS), is an approximately constant luminosity region in the ETD.
It is populated by ILOTs of different kinds: Intermediate-Luminous
Red Transients (ILRT), Luminous Red Novae (LRN) and LBV giant
eruptions, aka SN impostors. It also includes predicted BD-planets
merger-bursts (\citealt{Bear2011}). The upper limit of the OTS is
constrained by equation (\ref{eq:Lmax}). The green line represents
nova models computed using luminosity and duration from
\cite{dellaValleLivio1995}. Nova models from \cite{Yaron2005} are
marked with red crosses, and models from \cite{Sharaetal2010} are
represented with diamonds. The total energy does not include the
energy which is deposited in lifting the envelope that does not
escape from the star. For ILOTs that have model to estimate the
gravitational energy released by the accreted mass (the available
energy), it is marked by a black asterisk. PNe that might have
been created in ILOT events are marked with gray horizontal lines,
indicating the uncertainty in their ILOT event durations
(\citealt{SokerKashi2012}). }} \label{fig:ETD}
\end{figure*}

A panoramic way to examine and compare ILOTs and other transient
events is by using the Energy Time Diagram (ETD), as presented in
Fig. \ref{fig:ETD} (For a diagram showing peak luminosity rather
than total energy against the duration of the event, see, e.g.,
\citealt{Kulkarnietal2007} and \citealt{Kasliwal2013}.)
The diagram shows the total energy of the
transient as a function of its duration. Both the energy and the
duration are not easy to define. The total energy includes both
the integrated luminosity (radiated energy), and the kinetic
energy that went to the ejected material. The radiated energy is
calculated by integration of the the light curve in the optical
bands. The bolometric radiated energy is then calculated using a
bolometric correction that is not always accurate. The kinetic
energy is not easy to derive from observations, and may be a few
times larger than the radiated energy. It results in a range of
possible energies, that is presented by an error bar in the ETD.

In some cases the energy that is missed by observations, i.e., the
total available energy minus that emitted in the optical band, can
be large. This is the case for V838~Mon, where the energy required
to inflate the envelope, marked by the blue filled circle in Fig.
\ref{fig:ETD}, is more than $10$ times larger than that carried by
radiation (\citealt{TylendaSoker2006}). To account for the total
\emph{available} energy we calculate and present the gravitational
energy released by the accreted mass, as described in eq. (5)
of \cite{KashiSoker2010b}. This energy is marked by a black asterisk in the ETD.
The advantage of doing so is that the total available
energy better reflects the physical processes behind the
transient, as according to our model these transient events are
powered by gravitational energy. However, the total energy is
estimated using a detailed model, and those models were derived
for only a fraction of all transient events.

The ILOT time scale is defined typically as the time from its peak luminosity until it reaches its
pre-ILOT luminosity, before the rise. (Using a criterion like a decrease by 3 magnitudes
will not change much the graph.) Ideally it is defined for the
bolometric luminosity, and as observations are limited, the
waveband that contains most of the energy is used.

The ILOTs are found to lie on a wide stripe in the ETD -- the Optical
Transients Stripe (OTS). The stripe is not just an accidental
position, and has its roots in accretion physics, as we
demonstrate in section \ref{subsec:PhysicsIlots} below. Being a transient object
positioned on the OTS is one of the shared characteristics of
ILOTs, and in most cases it suggests that the transient is powered
by gravitation energy, and belongs to the ILOT group.

In previous papers we reported different ILOTs that lay on the OTS
(\citealt{KashiSoker2010b, Bear2011, SokerKashi2011, SokerKashi2012}).
The ETD presented in Fig. \ref{fig:ETD} is updated with a
number of recent ILOTs, as listed below.
\begin{enumerate}
\item \textbf{SN~2009ip}. An LBV that suffered several eruptions
before (e.g., \citealt{Berger2009a, Maza2009, Drake2012,
Levesque2014, Mauerhan2013, Pastorello2012}) its last one in 2012
(\citealt{Marguttietal2014, Martinetal2015}), that might have been a
real supernova explosion (\citealt{Smithetal2014}). There is also a
possibility that the last 2012 eruption was an exceptionally
strong giant eruption as well (\citealt{SokerKashi2013,
Kashietal2013}). The calculated location of the sub-eruptions in
2011 and the two in 2012 (2012a and 2012b), as well as the total
three years activity (SN~2009ip Tot), are marked on Fig.
\ref{fig:ETD}. It has been assumed that the total energy is twice
the radiated energy for all eruptions of SN 2009ip.
\item \textbf{SN~2010mc}. This supernova had a pre-explosion
outburst about a month before explosion (\citealt{Ofeketal2013}).
\cite{Soker2013b} proposed that the pre-explosion outburst was
powered by accretion onto a main sequence companion.
\item \textbf{R71 (HDE 269006)}. A giant eruption of an LBV.
Energy is calculated based on \cite{Mehneretal2013}. Late time
data is missing so the location may be somewhat more to the upper
right (longer duration and more energy), but certainly within the
OTS.
\item \textbf{CK Vul (Nova Vul 1670)}. An historic eruption
showing three peaks between 1670 and 1672 (\citealt{Sharaetal1985}).
\cite{Hajduketal2013} and \cite{Kaminskietak2015} suggested it
might have been an ILOT. We used historic data to position it on
the ETD. Observations of a bipolar nebula (\citealt{Hajduketal2007})
suggest the activity of jets.
\item \textbf{OGLE-2002-BLG-360}. \cite{Tylendaetal2013} claim it
to be an ILOT (an LRN).
\item \textbf{SN2007sv}. This seems to be a SN impostor according
to the analysis of \cite{Tartagliaetal2015}. We find it to be at
the center of the OTS. The total radiated energy is more than an
order of magnitude below that of typical SNe. Since more than $7$
years have passed since the outburst (ILOT event) and no SN
explosion took place, this ILOT is not excited by oxygen burning
in the core of a massive star. Its location on the OTS shows it is
one of the short duration and low energy SN impostors.
\item \textbf{M31LRN~2015}. A recent LRN (\citealt{Williamsetal2015}).
\item \textbf{SN~Hunt248}. This SN Impostor experienced an eruption with 3 peaks and returned to its pre eruption luminosity
in about 230 days  (\citealt{Kankareetal2015} and references therein).
Its position in the ETD is close to that of the first 2012 peak of SN~20009ip.
\end{enumerate}

\subsection{The high-accretion-powered ILOT (HAPI) model}
 \label{subsec:PhysicsIlots}

The most significant property ILOTs share according to the HAPI
model is that the energy source is gravitational. A high
accretion rate $\dot M_a$ onto a MS or a slightly evolved off-MS
star accounts for the high luminosity. We here explore some of the
implications of the HAPI model.

Let $M_a$ and $R_a$ be the mass and radius of the star accreting
the mass, respectively. Star `$b$' is the one that supplies the
mass to the accretion; it is possibly a MS star that is completely
destroyed, as in V838~Mon and V1309~Sco, or alternatively an evolved star in an
unstable phase of evolution that loses a huge amount of mass, as
in the GE of $\eta$ Car. The average total
gravitational power is the average accretion rate multiplied by the
potential well of the accreting star
\begin{equation}
L_G=\frac{G M_a \dot{M_a}}{R_a}.
\label{eq:L}
\end{equation}
The accreted mass will plausibly form an accretion disk or an
accretion belt. The accretion time ought to be longer than the
viscous time scale for the accreted mass to lose its angular
momentum. The viscous timescale is
\begin{equation}
\begin{split}
t_{\rm{visc}} &\simeq \frac{R_a^2}{\nu} \simeq 73
\left(\frac{\alpha}{0.1}\right)^{-1}
\left(\frac{H/R_a}{0.1}\right)^{-1}
\left(\frac{C_s/v_\phi}{0.1}\right)^{-1}\\
   &
   \times \left(\frac{R_a}{5~R_{\odot}}\right)^{3/2}
\left(\frac{M_a}{8~M_{\odot}}\right)^{-1/2} \days, 
\end{split}
\label{eq:tvisc1}
\end{equation}
where $\nu=\alpha ~C_s H$ is the viscosity of the disk, $H$ is the
thickness of the disk, $C_s$ is the sound speed, $\alpha$ is the
disk viscosity parameter, and $v_\phi$ is the Keplerian velocity.
We scale $M_a$ and $R_a$ in equation (\ref{eq:tvisc1}) to confirm with
to the parameters of V838~Mon (\citealt{Tylendaetal2005}). For these parameters,
the ratio of viscous to Keplerian timescale is $\chi \equiv
t_{\rm{visc}}/t_K \simeq 160$.

The accreted mass is determined by the details of the binary
interaction process, and varies for different objects. We scale it
by $M_{\rm{acc}} = \eta_a M_a$. Based on the modeled systems
(V838~Mon, V~1309~Sco, $\eta$ Car) this mass fraction is $\eta_a
\lesssim 0.1$ with a large variation. The value of $\eta_a
\lesssim 0.1$ can be understood as follows. If the MS star
collides with a star and tidally disrupts it, as in the model for
V838~Mon (\citealt{TylendaSoker2006, SokerTylenda2006}), the
destructed star is likely be less massive than the accretor
$M_{\rm{acc}} \lesssim M_b \lesssim 0.3 M_a$. In another possible
case an evolved star loses a huge amount of mass, but the accretor
gains only a small fraction of the ejected mass, as in the GE of
$\eta$ Car.

The viscous time scale gives an upper limit on the accretion rate
\begin{equation}
\begin{split}
\dot{M_a} < \frac{\eta_a M_a}{t_{\rm{visc}}} &\simeq 4
\left(\frac{\eta_a}{0.1}\right) \left(\frac{\alpha}{0.1}\right)
\left(\frac{H/R_a}{0.1}\right)
\left(\frac{C_s/v_\phi}{0.1}\right)\\
  &
  \times \left(\frac{R_a}{5~R_{\odot}}\right)^{-3/2}
\left(\frac{M_a}{8~M_{\odot}}\right)^{3/2}~M_{\odot} \yr^{-1}. 
\end{split}
\label{eq:dotM}
\end{equation}
The maximum gravitational power is therefore
\begin{equation}
\begin{split}
L_G < L_{\rm{max}} = \frac{GM_a\dot{M_a}}{R_a}& \simeq 7.7 \times
10^{41} \left(\frac{\eta_a}{0.1}\right)
\left(\frac{\chi}{160}\right)^{-1}\\
&
 \times \left(\frac{R_a}{5~R_{\odot}}\right)^{-5/2}
\left(\frac{M_a}{8~M_{\odot}}\right)^{5/2} \erg ~\rm{s^{-1}},
\end{split}
\label{eq:Lmax}
\end{equation}
where we replaced the parameters of the viscous time scale with
the ratio of viscous to Keplerian time $\chi$. Eq.
(\ref{eq:Lmax}) determines the upper bound on the OTS in the ETD. The
upper bound might be crossed if the accretion efficiency $\eta$ is
higher and/or the stellar parameters of the accreting star are
different. For most of the ILOTs the accretion efficiency is
lower, hence they are located below this line, giving rise to the
relatively large width of the OTS. The uncertainty in $\eta_a$
is large and may be even above unity, but only in extreme cases.
Therefore, we seldom expect to find objects slightly above the upper limit.

\subsection{Implications of Accretion Powered Events}
\label{subsec:implications}

Adopting the premise that ILOTs are powered by high mass accretion
rate, the HAPI model, a great deal can be learned about them.
We bring three examples.

The stellar masses of the two stars composing the $\eta$ Car
system were constrained using that model. The GE
of $\eta$ Car was initiated by close interaction with the
eccentric companion close to periastron passages
(\citealt{KashiSoker2010a}). \cite{Damineli1996} tried to support the
existence of a periodical variation using the historical data, but claimed it
cannot be done due to poor confidence on data.
Our calculation worked in an opposite way -- we used the available historical light curve to develop
a model and calculate the stellar and orbital parameters.

According to the model of \cite{KashiSoker2010a}, tidal forces of
the companion disrupt the LBV that was in an unstable state,
having an Eddington factor close to $1$. According to the HAPI
model the luminosity peaks of the GE resulted from accretion onto
the companion close to periastron passages
(\citealt{KashiSoker2010a}). Those peaks are $\sim 5.1$ years apart,
while the present orbital period is $\sim5.54$ years, suggesting
that during the GE in the 1840's, the orbital period was shorter.
Changes in orbital period came most likely from mass loss by the
erupting system (enlarges the period), and accretion from the more
massive LBV to the companion (shortens the period).
\cite{KashiSoker2010a} found that the only way for the HAPI
mechanism to work within the orbital changes constraints, requires
the two stars to have a significantly larger masses than initially
thought. Instead of a binary system composed of an LBV with a
mass of $M_1 \approx 120~M_\odot$ and a companion of $M_2 \approx
30~M_\odot$, the masses should be in the range  $M_1 \approx 170$
-- $200~M_\odot$ and $M_2 \approx 80~M_\odot$. Based on the
results of \cite{Figeretal1998} a zero-age main sequence (ZAMS)
star with an initial mass of $M_{\rm ZAMS} \approx 230~M_\odot$ is
required to explain the present luminosity of the primary of
$\eta$ Car. The massive primary star deduced here is expected to
results from a somewhat more massive ZAMS star, compatible with
$M_{\rm ZAMS} \approx 230~M_\odot$. This cannot be said on a
present primary mass of $120~M_\odot$.

The accretion of the mass ejected from the primary onto the
secondary formed a disk and launched jets. These jets had two
effects: (1) They shaped the Homunculus as a bipolar nebula; (2)
They may have cleared some of the mass in the vicinity of the
companion and by that reduced the amount of available gas for
further accretion, through the JFM discussed in section
\ref{sec:JFM}.

A similar idea was applied to the LBV star P~Cyg. There is no
observed companion to P~Cyg. Adopting the HAPI model and analyzing
the peaks in its historical light-curve, \cite{Kashi2010}
concluded that a $3$--$6~M_\odot$ binary companion in an eccentric
orbit most likely does exist. The periastron passages of the
companion caused the LBV to lose mass, part of which was accreted
by the companion, releasing energy. The whole process modified
the orbital parameters, resulting in a present period of $\sim 7$
years. The nebula around P~Cyg has bi-axial features that may have
been a result of relatively weak jets launched from the proposed
companion during the outbursts. If such a companion is not found,
it is possible that is was swallowed by the LBV star.

\cite{SokerKashi2012} proposed that a number of bipolar PNe may
have undergone ILOT events. Some PNe, such as NGC~6302 and the
pre-PNe OH231.8+4.2, M1-92, and IRAS~22036+5306 have bipolar
features that were formed in a rapid mass ejection event (e.g.,
\citealt{Meaburn2008, Szyszka2011}). The formation of a bipolar
nebula in an event lasting weeks to years is a prediction of the
HAPI model. The interaction with a MS companion causes the AGB (or
extreme AGB) to lose a huge amount of mass. Part of this mass is
accreted by the MS companion through an accretion disk. The disk
launches two jets that form the lobes. The process is accompanied
by high luminosity that makes the event an ILOT.

\section{THE NEGATIVE JET FEEDBACK MECHANISM}
 \label{sec:JFM}

Consider a flow structure where a large gas reservoir supplies
mass to an accretion disk that launches jets. If these jets are
not too narrow, they can interact with the reservoir and expel and
heat the mass residing there. This heating and/or ejection of
reservoir gas reduces the mass supply rate to the accretion disk,
hence lowering the jets' power, and even shutting-off jet formation
completely.
Jets can affect the ambient gas very efficiently by getting shocked to high temperature and inflating hot bubbles
(e.g., \citealt{HillelSoker2016} and references therein). The bubble has a larger cross section to push and expel ambient gas,
also in directions away from the direction of the jet.
The inflation of the bubble is accompanied by the formation of many vortices that mix part of the hot gas
with the cooler ambient gas (e.g., \citealt{HillelSoker2016}).
This results in an efficient heating of the ambient gas.
This entire process can convert kinetic energy to thermal energy and then to radiation.

This negative jet feedback mechanism (JFM) was
studied for a variety of astrophysical objects in the literature.
These systems are listed in Table 1.
The table does not include planetary nebulae (PNe), as for them the
interaction of the jets is on very large scales and it does not set a feedback.
Only if the shaping takes place close to the binary system, like if the nebula
is expelled during a GEE, that might also involved an ILOT (see \citealt{Soker2016}),
feedback occurs. It is listed here under GEE.
We compare the size and the gravitational potential of the
accreting body in each type of objects with those of the reservoir
of gas, $R_a$, $\Phi_a$, $R_{\rm res}$, and $\Phi_{\rm res}$,
respectively.
\begin{table*}[!t]
\renewcommand{\arraystretch}{1.1}
\begin{center}
\caption{Comparing Feedback Properties in Astrophysical Objects.}
\vspace{0.5cm}
\begin{tabular} {l c c c c c c }
\hline  
{Object type}  & Accreting   & $R_a$          & {$\Phi_a$}          & {$R_{\rm res}$} & {$\Phi_{\rm res}$}  & {$\Phi_a/\Phi_{\rm res}$}        \\
               & object      &  {(cm)}        & {$(\km \s^{-1})^2$} & {(cm)}          & {$(\km \s^{-1})^2$} &                                  \\
\hline  
\hline  

Galaxy            & SMBH        & $10^{11-14}$   &  $ c^2 $             &   $10^{20-23}$  &    $(300)^2$       &  $\approx 10^6$               \\
formation         &             &                &                      &                 &                    &                \vspace{0.2cm} \\

Cluster           & SMBH        & $10^{13-16}$   &  $ c^2 $             &   $10^{21-24}$  &    $(1\,000)^2$    &  $\approx 10^5$               \\
cooling flows     &             &                &                      &                 &                    &                \vspace{0.2cm} \\

Core collapse     &  NS         & $10^{6}$       &  $(100\,000)^2$      &   $10^9$        &    $(10\,000)^2$   &  $\approx 100$                \\
supernovae        &             &                &                      &                 &                    &                \vspace{0.2cm} \\

GEE $^{[1]}$      & MS star     & $10^{11}$      &  $(500)^2$           &   $10^{13}$     &    $(30)^2$        &  $\approx 100$ \vspace{0.2cm} \\

CEE $^{[1],[2]}$  & MS star  & $10^{11}$         &  $(500)^2$           &   $10^{13}$     &    $(30)^2$        &  $\approx 100$ \vspace{0.2cm} \\

ILOTs $^{[2]}$    & MS star     & $10^{11-12}$   &  $(1\,000)^2$        &   $10^{11-13}$  &    $(30-1\,000)^2$ &  $\approx 3-100$              \\
\hline  
\end{tabular}
\label{tab:Table1}\\
\end{center}
\begin{flushleft}
\small Notes: \\
$R_a$ and $\Phi_a$ stand for the typical radius of the accreting
object, and the gravitational potential on its surface. $R_{\rm
res}$ and $\Phi_{\rm res}$ stand for the typical radius of the
reservoir of gas for accretion and the energy required to expel it
from the system. $c$ represents the speed of light.
\newline
Abbreviations: BH: black hole; CCSN: core-collapse supernova; CEE:
common envelope evolution; GEE: grazing-envelope evolution; NS:
neutron star; SMBH: super-massive BH.
\newline
 \small \footnotemark[1]{We refer here only to a giant primary star and a main sequence (MS) companion. See text for the case of a NS or a BH
 companion.}
 \newline
 \footnotemark[2]{While in the other objects jets are a crucial ingredient in the evolution, in the CEE and and in ILOTs in some cases jets do not occur, or play a small role.
 Furthermore, not all cases the JFM operates even if jets do exist.}
\end{flushleft}
\label{Table1}
\end{table*}

\cite{Sokeretal2013} made comparisons of the JFM operating during
galaxy formation, in cooling flows in clusters of galaxies, during
the common envelope evolution (CEE), in PNe, and in core collapse
supernovae (CCSNe). Boosted by the results of the Chandra and
XMM-Newton X-ray observatories, there have been many papers in the
last fifteen years on the operation of a JFM in galaxy formation
and cooling flow clusters (\citealt{McNamaraNulsen2012}; \citealt{Sokeretal2013} and references therein).
In recent years there have been suggestions
for the operation of a JFM in the explosion of massive stars in
core collapse supernovae (CCSNe), and during the CEE. The
jet-feedback mechanism for the explosion of all CCSNe is termed
the jittering-jets model (\citealt{Papish2011, Papish2012,
PapishSoker2014a, PapishSoker2014b, GilkisSoker2014,
GilkisSoker2015a, GilkisSoker2015b, Papishetal2016}).

Numerical simulations encountered difficulties in expelling the
envelope gas during the CEE, as a large fraction of the envelope
stays bound to the system even after a substantial spiraling-in of
the two stars (e.g., \citealt{SandquistTaam1998, Lombardi2006,
DeMarco2011, Passy2011, Passyetal2012a, RickerTaam2012}, but see \citealt{Nandezetal2015}).
These difficulties and the problems with the $\alpha_{\rm
CE}-$prescription (e.g., \citealt{Soker2013a}) lead to the
suggestion that in many cases envelope ejection is facilitated by
jets launched by the more compact companion (\citealt{Soker2004,
KashiSoker2011, Soker2013a, Soker2014}). \cite{ArmitageLivio2000}
and \cite{Chevalier2012} studied CE ejection by jets launched from
a neutron star (NS) companion but not as a general CE ejection process.

An efficient JFM might even prevent the formation of a CEE.
Instead, the compact companion spirals-in while grazing the
envelope of the giant star, in what is termed the grazing envelope
evolution (GEE; \citealt{Soker2015}). If the companion is on an
eccentric orbit and grazes the envelope, or even forms a temporary
common envelope, it can accrete at a high rate, releasing
gravitational energy that is observed as an ILOT. As such, the JFM
can operate in ILOTs, as we propose below. We note that
\cite{Ivanovaetal2013} mention that the ejection of the envelope
through a CCE can lead to an ILOT event, but they mention no jets.
 
Launching jets by a MS star in the GEE and the CEE requires that MS stars can accrete at high rates reaching $0.001$--$0.1 M_\odot \yr^{-1}$, and launch jets.
It seems that MS stars can indeed accrete mass at high rates if the jets that are launched by the accretion disk
or belt remove most of the accreted energy and angular momentum (\citealt{Shiberetal2016}).
As well, even if the specific angular momentum of the accreted gas is below the value needed to form a Keplerian accretion disk,
but not by much, the accretion belt formed around the MS star might launch jets (\citealt{SchreierSoker2016}).
An accretion belt is defined here as a sub-Keplerian accretion flow on the surface of the accreting body, with mass concentration closer to the equatorial plane. \cite{SchreierSoker2016} argue that jets might be blown from the polar regions of the accretion belt.

The effect of the JFM can be quantified by the ratio between the
gravitational potential energy of the mass accreted onto the
surface of the compact body, and that of the ambient gas in the
reservoir. If this ratio is large, as in galaxies and clusters of
galaxies, then a small amount of accreted mass can launch jets
that have a large impact on the ambient gas. We find that for many
ILOTs this ratio is smaller than in the other types of objects
listed in Table 1. This implies that the JFM in ILOTs is less
efficient that in most other cases listed in Table 1. A large
fraction of the mass available in the reservoir is accreted in a
typical ILOT event.

We note that in the case of ILOTs, jets can interact with gas
residing close to the accreting star, hence leading to the
operation of the JFM, or with gas residing further out that is not
part of the reservoir. The interaction with gas residing further
out, but in a still optically thick region, converts kinetic
energy to thermal energy, and then radiation. Namely, the
interaction can increase the ILOTs luminosity.

We turn now to discuss several settings of ILOTs powered by
accreting MS stars, and the different manifestation of the JFM in
each one of them. The three different scenarios differ mainly
in the time period during which the accretion process takes place,
and in the reservoir of the accreted mass onto the MS star.

(1) \emph{ An ILOT event at a periastron passage.}
In this scenario the accretion phase lasts for a fraction of the orbit.
The accreted gas originates in one side of the bloated atmosphere of a giant star.
For these, the JFM does not destroy the entire mass reservoir, and if the MS companion survives,
then the outburst might repeat itself. The best example is the GE of $\eta$ Car,
where there were at least two outbursts separated by $\sim5.2$ years of orbital period.
Both mass removal and mass accretion act to change both the eccentricity and the orbital separation,
but in opposite directions. Whether the eccentricity and the orbital separation increase or decrease
depends on the detail of the interaction, and on the mass loss rate far from periastron,
when accretion does not take place or has a small rate (see \citealt{KashiSoker2010b}).
In the GE of $\eta$ Car, according to the HAPI model, the companion accreted
mass mainly at periastron passages. The distance between the two
stars at periastron passages was $a_p \approx 300 R_\odot$. The
companion could have entered the extended and tidally distorted
envelope of the primary LBV star, and then get out. This may be
termed GEE. Most of the accreted mass came from the equatorial
plane. Some mass though, could have been accreted from all
direction including from above and below the companion, at a
typical distance from the secondary star of $\approx
50$--$100~R_\odot$. The radius of the secondary star is $R_a
\approx 20~R_\odot$. When the mass of the primary star is
included, we find $\Phi_a/\Phi_{\rm res} \approx 3$. This
relatively low value and the small fraction of reservoir gas
residing along the polar directions make the JFM less efficient
than in most objects listed in Table 1. Still, we hold that the
JFM can operate in giant eruptions of very massive stars.

(2) \emph{The JFM in a fall back gas.}
In this scenario the JFM takes place a long time, relative to the dynamical time,
after the outburst was triggered. The accreted gas is a fall back gas from the violent event,
e.g., a merger process. This JFM is expected to operate once after one violent event.
As not much mass falls back, the JFM here will not change much the binary parameters.
The estimated ejected mass in the 1988 outburst of M31~RV was $\approx 0.001$--$0.1~M_\odot$,
and with an expansion velocity of $\approx 100$--$500~\km~\s^{-1}$
(\citealt{Mouldetal1990}). Based on some similarities with V838 Mon,
\cite{SokerTylenda2003} suggested that M31 RV eruption was caused
by merger-burst event. In \cite{KashiSoker2010b} we raised the
possibility that a merging processes might be replaced with a
rapid mass transfer episode in a binary system. We considered a
small ejected mass that quickly became optically thin. Part of the
ejected material might have not reached the escape velocity, and
fell back toward the star, or the binary system if the companion
survived the event. The inflated envelope and the gas that did not
reach the escape velocity fell back toward the star. Instead of a
long-lived inflated envelope as in the case of V838 Mon, we
argued, an accretion disk was formed after about a year. Because
of the high specific angular momentum in the binary system the
fall back gas formed an accretion disk around the accreting star.
The material closer to the center formed an accretion disk before
the outer parts of the envelope have collapsed.

Here we add that such a scenario can lead to the onset of the
JFM. The disk from the fall back gas might have launched jets. The
jets could have expelled some of the outer parts of the envelope
that did not form a disk yet, hence lowering the amount of gas
accreted to the center. This is a negative JFM process. It might
act in general cases of merger-bursts where a large envelope is
inflated, and in addition the gas that continues to be accreted
onto the central star has sufficient specific angular momentum to
form an accretion disk that launches two jets.
Namely, even if most of the mass in a red transient (a merger-burst)
is lost via the outer Lagrange point, e.g., \cite{Pejchaetal2015},
jets might still be formed. This can be after the merger process
as outlined above, or before the full merger when the denser star
accretes gas from the other star via an accretion disk.

The ratio $\Phi_a/\Phi_{\rm res}$ depends on the mass distribution
in the envelope that is formed after merger. Even if it is an
extended envelope, most of the reservoir mass might reside close
to the accreting star, resulting in $\Phi_a/\Phi_{\rm res} \approx
10$. As well, due to the high angular momentum of a merger event,
mass is concentrated toward the equatorial plane, hence less
affected by the polar jets. Over all, the JFM is not extremely
efficient in this setting. Still, it plays some role.

We note that in the merger simulation of \cite{Nandezetal2014} no
accretion disk was formed. We do not claim that jets are formed in
all merger-bursts, but in many cases they might form.

(3) \emph{Spiral-out grazing envelope evolution.}
In this scenario the accretion phase lasts for the entire orbit, and for many orbits.
The accreted gas originates from the outer parts of the giant's envelope and near its equator.
Mass removal can be significant, hence acting to increase orbital separation.
But as the JFM requires the MS companion to accrete mass from the envelope,
even if orbital separation increases, it will not be by much, as the companion
cannot get too far from the envelope. Basically, the companion might end at an orbital
separation of about the size of the giant primary star, $\approx 1 \AU$. 
The removal of gas by jets can have an indirect influence through dynamical
effects. This is known to occur for example during galaxy
formation. A rapid removal of baryonic mass by AGN activity
reduces the depth of the gravitational well, hence causing also
the dark matter to expand (e.g., \citealt{Martizzietal2013}).
This further reduces the gravitational
well, and hence acts to further lower the density of gas in the
center of the galaxy. This in turn acts to reduce accretion rate
to the central SMBH.

\begin{figure*}[!t]
    \centering
{\includegraphics*[scale=0.52]{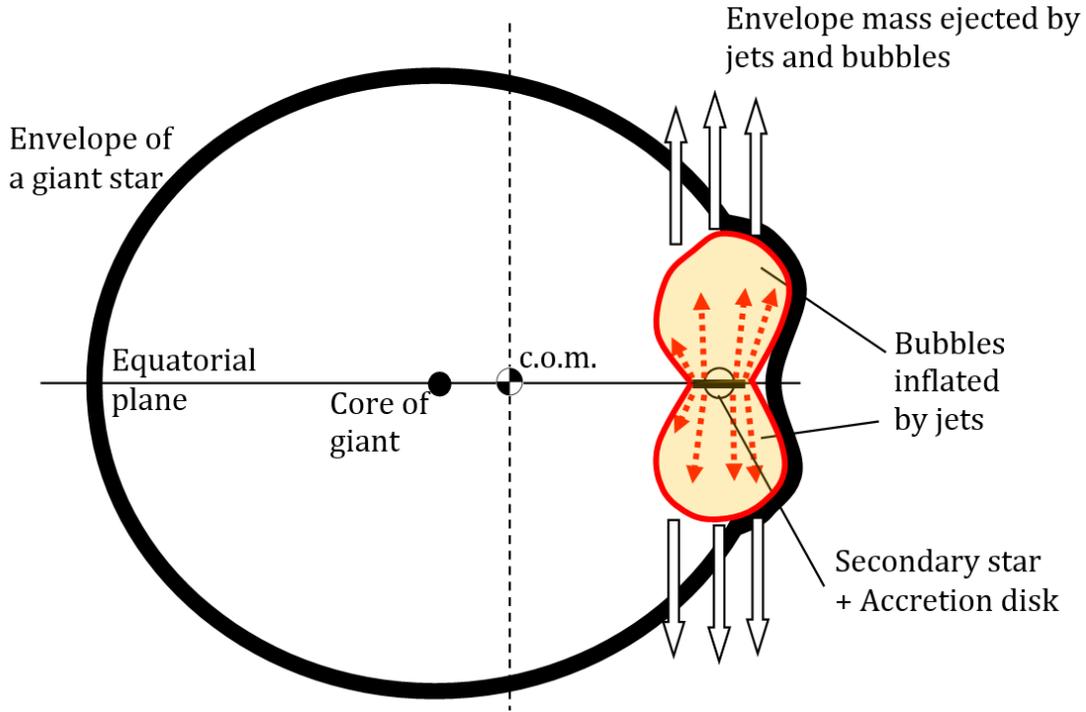}}
 \caption{A schematic flow structure of a grazing-envelope evolution (GEE).
The companion accretes mass through an accretion disk and launches
two jets in the outskirts of the giant envelope. The jets inflate
bubbles and efficiently remove most of envelope mass residing
above and below the secondary orbit. A large fraction of this mass
is located not far from the axis through the center of mass,
depicted by a dashed line perpendicular to the orbital plane, and
hence posses a low value of specific angular momentum.}
  \label{fig:GEE}
\end{figure*}

Mass loss by either star in a binary system and mass transfer
between them change the orbital parameters. The operation of the
JFM in removing gas from a giant envelope (the primary) can have a dynamical
effect. The jets remove mass above and below the trajectory of the
accreting compact star through the giant envelope, as depicted in
Fig. \ref{fig:GEE}. As the center of mass is between the two
stars, in the case that the secondary star is not much lighter than
the primary star, the removed mass is closer to an axis through
the center of mass and perpendicular to the equatorial plane
(depicted by a dashed line) than most of the rest of the giant's
surface. This implies that the removed envelope mass has a lower
value of specific angular momentum than the average of the giant's
surface.
Since in a strong binary interaction it is expected that the mass loss
will be concentrated to the equatorial plane, we find that the
JFM removes mass with a lower specific angular momentum than that
from an equatorial mass loss from the primary star. Therefore, the
mass removal by jets has a less pronounced effect on reducing the
orbital separation. In addition, the energy used to expel the
envelope mass comes from the mass accretion onto the companion,
and not from the orbital energy. These two properties, of a
relatively low specific angular momentum of the removed gas and
using the accretion energy to remove gas, do not act strongly to
reduce the orbital separation.
Therefore, the mass loss process in the JFM acts to increase the orbital separation.
On the other hand, if the accreting secondary
star is much lighter than the primary giant star, mass transfer
acts to reduce the orbital separation.

If the accretion rate started at a high rate and later decreased,
there may be enough gas in the disk to lunch a jet. The mass
removal by jet can supersede the decreasing accretion rate,
deplete the reservoir and consequently stop the mass transfer.
This would cause an increase of the orbital period. If the
binary has an eccentric orbit, eccentricity can increase as well.
This spiral-out evolution resulting from the JFM operating in
binary stars will be studied in detail in a forthcoming paper.

\section{SUMMARY}
\label{sec:summary}

We explored some aspects of the physics of ILOTs under the
assumption that they are powered by high mass accretion rate on to
a MS, or slightly evolved off the MS star.
In this high-accretion-power ILOTs (HAPI) model one star in a
binary system accretes mass from a companion. The companion can
survive or be completely destroyed in a merger-burst event.

In section \ref{subsec:ETD} we defined ILOTs as the general group
of gap objects, between novae and supernovae, that are not type
.Ia SNe or related objects. We also classified ILOTs into three
subgroups according to the mass transfer process. We updated the
ETD where ILOTs occupy a stripe, termed the OTS (Fig.
\ref{fig:ETD}). The upper boundary of the OTS can be accounted for
by accretion via an accretion disk around a MS star. This upper
boundary was derived in section \ref{subsec:PhysicsIlots} (eq.
\ref{eq:Lmax}). Some implications of the HAPI model were discussed
in section \ref{subsec:implications}. In particular, we reiterate
an earlier result that the masses of the two stars composing the
LBV binary system $\eta$ Car are much more massive than what is
usually assumed based only on the Eddington luminosity limit.

Our most significant new claim is that in some cases ILOTs are
regulated to some degree by a negative feedback mechanism mediated
by jets. In this JFM the jets expel some mass from the ambient gas
reservoir of the accretion disk. This in turn reduces and/or
shortens the mass accretion rate. In Table 1 we compared some
properties of the JFM of ILOTs with other types of astrophysical
objects. In section \ref{sec:JFM} we discussed three examples of
the possible operation of the JFM in ILOTs.

The examples listed in section \ref{sec:JFM} show that the JFM in
ILOTs is less efficient than in most other types objects listed in
Table 1 for two reasons. ($i$) The value of the ratio of
gravitational potential on the mass accreting body to that in the
reservoir $\Phi_a/\Phi_{\rm res}$, is typically lower than in
other types of objects. This implies that more jets' mass is
required to remove a given amount of reservoir gas. ($ii$) Because
the interaction occurs relatively close to the accreting objects,
the high value of the specific angular momentum of the accreted
gas implies that it tends to flow closer to the equatorial plane.
As the jets are lunched along the polar directions, they interact
with a smaller fraction of the ambient reservoir gas than in a
spherical distribution.

As more ILOTs are being discovered we continue to populate the ETD\footnote{An updated version of the ETD is available at \url{http://physics.technion.ac.il/~ILOT/}} and learn more about them.
Further work is required to quantify the outcome of the JFM.
Numerical simulations will be the next step. These can show that the jets actually casue the depletion of the
mass reservoir.
Most relevant is the study of the newly proposed GEE, that requires 3D hydrodynamical numerical simulations.

\vspace*{0.5cm}
\begin{acknowledgements}
We thank an anonymous referee for helpful suggestions.
AK acknowledges support provided by National Science Foundation through grant AST-1109394.
AK would like to thank the Technion for generous hospitality, where part of this work was done.
\end{acknowledgements}

{}

\label{lastpage}
\end{document}